\documentclass[%
 reprint,
superscriptaddress,
 amsmath,amssymb,
 aps,
]{revtex4-2}

\usepackage{graphicx}
\usepackage{dcolumn}
\usepackage{bm}
\usepackage{subfigure}
\usepackage[dvipsnames]{xcolor}

\begin{document}


\title{Attosecond Entangled Photons from Two-Photon Decay of Metastable Atoms: A Source for Attosecond Experiments and Beyond}



\author{Yimeng Wang}
\email{wang3607@purdue.edu}
\author{Siddhant Pandey}%
\affiliation{Department of Physics and Astronomy, Purdue University, West Lafayette, Indiana 47907}
\author{Chris H. Greene}
\author{Niranjan Shivaram}
\email{niranjan@purdue.edu}
\affiliation{Department of Physics and Astronomy, Purdue University, West Lafayette, Indiana 47907}
\affiliation{Purdue Quantum Science and Engineering Institute, Purdue University, West Lafayette, Indiana 47907}%

\date{\today}

\begin{abstract}
We propose the generation of attosecond entangled bi-photons in the extreme-ultraviolet regime by two-photon decay of a metastable atomic state as a source similar to spontaneous parametric down-conversion photons. The 1s2s $^1S_0$ metastable state in helium decays to the ground state by emission of two energy-time entangled photons with a photon bandwidth equal to the total energy spacing of 20.62 eV. This results in a pair correlation time in the attosecond regime making these entangled photons a highly suitable source for attosecond pump-probe experiments. The bi-photon generation rate from a direct four photon excitation of helium at 240 nm is calculated and used to assess some feasible schemes to generate these bi-photons. Possible applications of entangled bi-photons in attosecond time scale experiments, and a discussion of their potential to reach the zeptosecond regime are presented. 
\end{abstract}

\maketitle


Quantum entanglement is a fascinating quantum phenomenon that has no classical analog \cite{horodecki2009}. Entanglement is at the heart of quantum information science, quantum sensing, quantum enhanced imaging and spectroscopy, and other emerging quantum technologies. Entanglement of photons has particularly played an important role in many areas of basic and applied research that leverage the quantum advantage. For example, entangled photons have been used in nonlinear spectroscopy \cite{lee2006,Schlawin2018,Saleh1998} which goes beyond the time-frequency uncertainty limit \cite{Strekalov2005,Pfeifer1995, dorfman2016}; Moreover, a linear (rather than quadratic) scaling of two-photon absorption rate versus intensity is observed with entangled photons \cite{Perrina1998,Fei1997,Schlawin2018,dorfman2016}, which enhances the process at low intensities. As a light source, entangled photons can collectively excite uncoupled atoms \cite{Muthukrishnan2004,Richter2011}, and lead to entanglement-induced two-photon transparency \cite{Fei1997}, which cannot be obtained by a classical laser source.  

Typical sources of entangled photons use the process of spontaneous parametric down-conversion (SPDC) in nonlinear crystals in the visible and infra-red region of the spectrum \cite{kwiat1995}. These sources generate energy-time entangled photons with correlation times on the femtosecond time scale which has been only recently directly measured \cite{maclean2018}. SPDC has also been demonstrated in the hard X-ray regime where the correlation times are expected to be attoseconds or smaller \cite{sofer2019}.  Recent experiments using nanophotonic chips for SPDC have demonstrated entangled photon generation with broad bandwidth of 100 THz (0.41 $eV$) and a high generation efficiency of 13 GHz/mW \cite{javid2021}. Here we propose a method to generate entangled photon pairs in the extreme-ultraviolet (XUV) regime with ultra-broad energy bandwidth ($>20$ $eV$) large enough to create correlation times on the attosecond scale.  


It is well known that the 1s2s $^1S_0$ metastable state of helium atom, its isoelectronic ions and the 2s $^2S_{1/2}$ metastable state of the helium ion decay predominantly by two-photon emission \cite{bondy2020,lipeles1965, drake1969, prior1972, mokler2004}. The emitted photons are energy-time entangled with a correlation time related to the energy spacing between the 2s and 1s levels which is 20.62 $eV$ and 40.81 $eV$ for the helium atom and ion respectively. This large energy bandwidth of the emitted entangled photons corresponds to correlation times in the attosecond domain, thus opening up the possibility of attosecond time scale pump-probe experiments using these photons.    

\begin{figure}[t]
\centering
    \begin{subfigure}
        \centering
        \includegraphics[width=0.34 \textwidth]{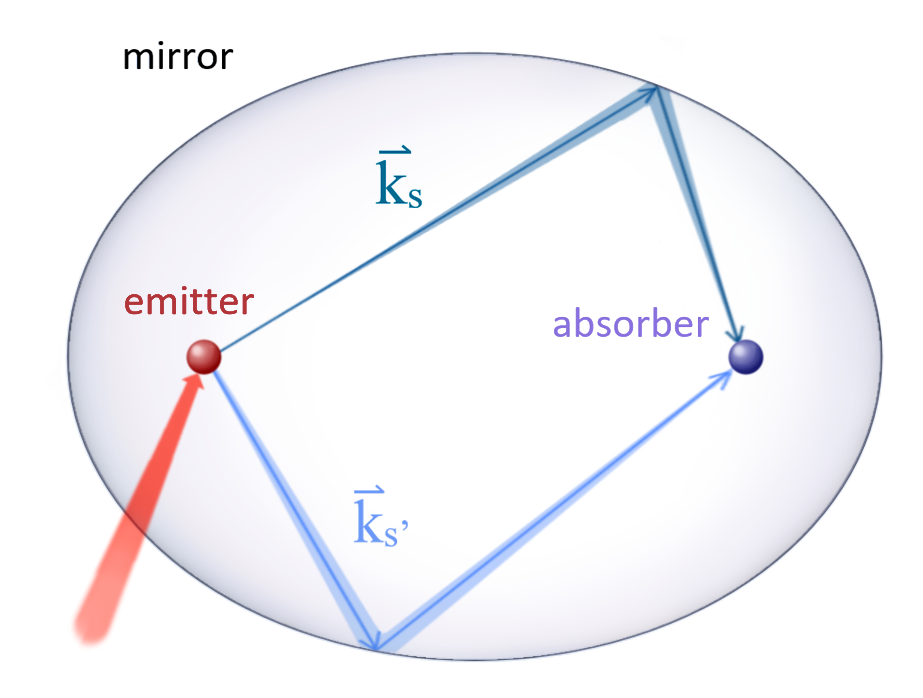}
        \caption{\label{fig:ell} A schematic diagram of entangled-photon generation and absorption in a spheroid cavity. The emission and absorption atoms are placed at the two foci of the spheroid, the photons are reflected by the boundary of the cavity, and propagate along equal paths to reach the absorber. The shape of the cavity will influence the rate of this process, by a geometry factor as discussed in the Supplementary Material.}
    \end{subfigure}
\end{figure}

We first consider a gedanken experimental set-up in which we have a spheroidal cavity, with two helium atoms placed at its two foci. One of the atoms is prepared in the $1s2s$ $(^1S_0)$ excited state, which is used as an emitter (atom 1), and another atom is in $1s^2$ $(^1S_0)$ ground state, which is used as an absorber (atom 2), as shown in Fig. \ref{fig:ell}. Atom 1 decays to the $1s^2$ ground state by a simultaneous emission of two photons, according to selection rules. This decay channel dominates over the magnetic dipole transition to the 1s2s $^3S_1$ state. Since the long-lived metastable $1s2s$ $(^1S_0)$ state has a long lifetime of $\tau=0.0197$ $sec$ \cite{vandyck1971}, and since the energy gap between $1s2s$ $(^1S_0)$ and $1s^2$ state is 20.62 $eV$, the two emitted photons have both a good correlation in frequency and a narrow window in emission time difference; energy-time uncertainty implies that they are good sources of entanglement. The bi-photons should also be correlated in angular momentum, according to the angular momentum conservation rule. However, we do not address that aspect, since inside a spheroid cavity the entangled photon-pair will be collected at the absorber with equal distance optical path, irrespective of their angular distribution or momenta. In treating this process, we assume that: 1. The cavity is large enough, that no quantization of photon frequencies or Purcell effect is relevant. 2. Both atoms are deeply trapped, so no recoil effects can be observed. 3. The mirror of the cavity is $100\%$ reflective to all the frequencies, so no energy loss occurs during reflection of the photons. Further, it is noted that the emitted bi-photons are also polarization entangled but we do not discuss polarization entanglement in this letter. All possible polarization configurations are considered in our calculation.



Inside the cavity, there are three stages of photoelectric processes : the population inversion of atom 1, the spontaneous two-photon emission of atom 1, and the photoabsorption of atom 2. In the first stage, we prepare the singlet $1s2s$ state using four photon absorption with each photon having energy $\hbar\omega_0=5.155$ $eV$ ($240.54$ $nm$). 
With a monochromatic incident electric field $\mathcal{E}_0\hat{\epsilon}_0\cos{(\omega_0 t)}$, the four-photon excitation amplitude is,
\begin{small}
\begin{equation}
\label{cor-1}
\begin{split}
     &C_{exc}(t)=\left(\frac{e\mathcal{E}_0}{2\hbar}\right)^4 \frac{e^{i(\Delta_{eg}-4\omega_0)t}-1}{\Delta_{eg}-4\omega_0}
     D^{(4)}_{eg}\\
     &D^{(4)}_{eg}= \sum_{j_1,j_2,j_3}\frac{\langle e|
    \hat{\epsilon}_0\cdot\vec{r}|j_3 \rangle\langle j_3|\hat{\epsilon}_0\cdot\vec{r}|j_2 \rangle
     \langle j_2|\hat{\epsilon}_0\cdot\vec{r}|j_1 \rangle\langle j_1|\hat{\epsilon}_0\cdot\vec{r}|g \rangle}
     {(\Delta_{j_3g}-3\omega_0)(\Delta_{j_2g}-2\omega_0)(\Delta_{j_1g}-\omega_0)}
\end{split}
\end{equation}
\end{small}
where $|g\rangle$ is the $1s^2$ ground and the initial state, $|e\rangle$ is the $1s2s$ excited and the final state, $|j_{1,2,3}\rangle$ are the intermediate states. 
Since 
$$\lim\limits_{t\to\infty}\frac{e^{i(\Delta_{eg}-4\omega_0)t}-1}{\Delta_{eg}-4\omega_0}=-\mathcal{P}(\frac{1}{\Delta_{eg}-4\omega_0})+i\pi\delta(\Delta_{eg}-4\omega_0) $$
The resulting unnormalized state following the excitation, which is also the initial state for the emission process, is: $|\gamma\rangle=i\pi\delta(\Delta_{eg}-4\omega_0) \left(\frac{e\mathcal{E}_0}{2\hbar}\right)^4 D^{(4)}_{eg} |e\rangle$. 

The photon-atom interaction for the second and third stages is: 
\begin{equation}
\begin{split}
        &V^{int}(t)=e \Vec{r}\cdot\vec{E}(t) \\
        &=e \Vec{r}\cdot\sum_s i \hat{\epsilon_s}\left(\frac{2\pi\hbar\omega_s}{V}\right)^{\frac{1}{2}}(a_s e^{i(\vec{k_s}\cdot\vec{r}-\omega_s t)}-a_s^{\dagger} e^{-i(\vec{k_s}\cdot\vec{r}-\omega_s t)}) \\
\end{split}
\end{equation}
Where $\vec{r}$ is the space vector of the electron, $\vec{E}$ is the electric field, and $V$ is the quantization volume, with the electric field generated by a single photon proportional to $1/\sqrt{V}$. The photon modes $s$ include the frequency $\omega_s$, propagation direction $\hat{k}_s$ and polarization direction $\hat{\epsilon}_{s}$. From a second-order perturbation analysis, the amplitude of the emission of two photons ($|\gamma\rangle\otimes|vac\rangle \rightarrow |g\rangle\otimes|1_s,1_{s^{\prime}}\rangle$) is, 
\begin{equation}
\label{cor-2}
\begin{split}
     C_{emi}^{(s,s^{\prime})}(t)&=-\frac{2\pi e^2}{V }\sqrt{\omega_s\omega_{s^{\prime}}} \frac{e^{i(\omega_s+\omega_{s^{\prime}}-\Delta_{eg})t}-1}{\omega_s+\omega_{s^{\prime}}-\Delta_{eg}}\\
     & \times\sum_j \frac{\langle g|\hat{\epsilon}_{s^{\prime}}\cdot\vec{r}|j \rangle\langle j|\hat{\epsilon}_s\cdot\vec{r}|\gamma \rangle}{\hbar(\omega_s-\Delta_{ej})}   \\
\end{split}
\end{equation}
$|j\rangle$ denotes the intermediate states for the emission process. $\Delta_{ej}$($\Delta_{eg}$) is the energy difference between the initial and intermediate (final) atomic state. 
From Eq. \ref{cor-2} we obtain the He $1s2s$ $(^1S_0)$ lifetime as $\tau=0.0197$ $sec$, which agrees with the experimental value \cite{vandyck1971}.


Since no singlet energy level exists between $E_i$ and $E_i+\Delta_{eg}$ for atom 2, the absorption process can only start after both photons have been emitted, with $\omega_s+\omega_{s^{\prime}}=\Delta_{eg}$. 
The modes of the photons are not detectable inside the cavity, therefore the entangled photon state can be obtained by summing over all the modes $(s,s^{\prime})$\cite{Shih_2003}:
\begin{equation}
\begin{split}
     |2ph\rangle&=\sum_{s,s^{\prime}}C_{emi}^{(s,s^{\prime})}(t\rightarrow\infty)|1_{s},1_{s^{\prime}}\rangle\\
\end{split}
\end{equation}

\begin{figure}
    \includegraphics[scale=0.53]{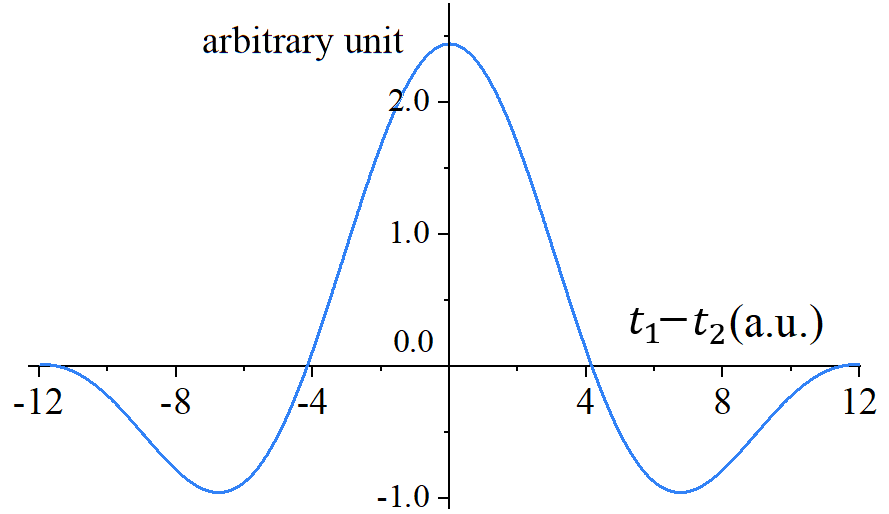}
    \caption{The photon correlation function $\langle vac|{E}(t_2){E}(t_1)|2ph\rangle$ (up to a constant factor) as a function of time difference $(t_2-t_1)$, which indicates the correlation time is around $1.93\times 10^{-16}$ $sec$.}
  \label{c3}
\end{figure}

Based on a second-order perturbation calculation, The entangled-photon absorption amplitude can be written as,
\begin{equation}
\label{cor-4}
\begin{split}
    &C_{abs}(t)
    = -\frac{e^2}{\hbar^2}\int_0 ^{t} dt_2 \int_{-\infty}^{t_2} dt_1 \sum_{m} e^{i(\Delta_{mi}t_1+\Delta_{fm}t_2)} \\
    & \times (\langle f|\otimes\langle vac|) \Vec{r}\cdot\vec{E}(t_2)|m\rangle\langle m|\Vec{r}\cdot\vec{E}(t_1) (|i\rangle\otimes|2ph\rangle)\\
\end{split}
\end{equation}
$|i\rangle$, $|m\rangle$ and $|f\rangle$ denotes the initial, intermediate, and final states for atom 2. $\vec{E}(t_{1,2})$ are the electric fields of the photons that are bounced back by the cavity (whose frequencies stay the same but propagation and polarization directions have changed), being absorbed at time $t_1$ and $t_2$. The evaluation of Eq. \ref{cor-4} depends on the shape of the cavity, and it turns out that the absorption process can be described by a rank-$0$ tensor, which is discussed in the Supplementary Material. 

The time correlation of the entangled photon pair can be found from $\langle vac|{E}(t_2){E}(t_1)|2ph\rangle$, which is proportional to the fourier transformation of the spectrum \cite{Titulaer1965,Perrina1998,Fei1997}, as
\begin{widetext}
\begin{equation}
\label{cor-5}
\begin{split}
     \langle vac|{E}(t_2){E}(t_1)|2ph\rangle
     &\propto 
     \int_{0}^{\Delta_{eg}} d\omega_s e^{i\omega_s (t_2-t_1)}[\omega_s(\Delta_{eg}-\omega_s)]^3
     \sum_{j}\left(\frac{\langle g|r|j \rangle\langle j|r|e \rangle}{\hbar(\omega_s-\Delta_{ej})}
     +\frac{\langle g|r|j \rangle\langle j|r|e \rangle}{\hbar(\Delta_{jg}-\omega_s)}\right)\\
\end{split}
\end{equation}
The right hand side of Eq. \ref{cor-5} is plotted in Fig.~\ref{c3}, versus the time difference between two absorption events of the two photons. The time scale between the two absorption events is around $\pm 4$ a.u. which gives a correlation time \cite{Strekalov2005} around $193$ attoseconds.


Finally, according to Eq.\ref{cor-1}, \ref{cor-2} and \ref{cor-4}, the rate for the excitation, emission and absorption where an entangled photon-pair is transferred coherently, is:
\begin{small}
\begin{equation}
\label{cor-7}
\begin{split}
    &R_{trans}=
    2\pi\delta(\Delta_{fi}-\Delta_{eg})
    \left| \frac{\Theta e^8 \mathcal{E}_0^4}{256 \hbar^6c^6} D^{(4)}_{eg}\delta(\Delta_{eg}-4\omega_0) \int d\omega_s [\omega_s(\Delta_{eg}-\omega_s)]^3
    \sum_{m}\frac{\langle f|r|m\rangle\langle m|r|i\rangle}{\omega_s-\Delta_{mi}} 
     \sum_{j}\left(\frac{\langle g|r|j \rangle\langle j|r|e \rangle}{\omega_s-\Delta_{ej}}
     +\frac{\langle g|r|j \rangle\langle j|r|e \rangle}{\Delta_{jg}-\omega_s}\right)\right|^2\\
\end{split}
\end{equation}
\end{small}

\end{widetext}

 
$\Theta$ is a geometry factor which is introduced in Eq. S3, whose values are shown in Figure S1 in the Supplementary Material. Especially, for a spherical cavity, $\Theta=\frac{64\pi^2}{27}$, $R_{trans}=1.91\times10^{-25} \mathcal{E}_0^8$ $a.u.$. The input beam flux is $J=\frac{c\mathcal{E}_0^2}{8\pi\hbar\omega_0}$. It is seen that the transition rate is proportional to $J^4$. The entangled-photon absorption rate is known to be proportional to the beam intensity (when the beam intensity is not very strong) \cite{Fei1997,lee2006,dorfman2016}, and our result can be regarded as a generalization of this linearity. Since our excitation process is a four-photon process, we can consider the four-photon flux as a whole, which is the input of the system, $J^{(4)}=J^4$. Therefore it is an expected outcome that $R_{trans}\propto J^{(4)}$. 

The above calculations assume a direct multiphoton excitation from $1s^2$ to $1s2s$. Since the $1s2s$ $^1S_0$ metastable state has a narrow linewidth of $\sim$ 50 Hz, ideally a multiphoton excitation to this state requires intense lasers with a linewidth smaller than 50 Hz at a wavelength of 240 nm. While multiphoton excitations of metastable states with narrow linewidth lasers have been previously demonstrated \cite{parthey2011}, achieving the required high intensities  with a narrow-band 240 nm laser is currently challenging. However, femtosecond lasers that can achieve peak intensities of $\sim 10^{14}$ $Wcm^{-2}$ are readily available. Using our calculations for the four-photon excitation amplitude with a monochromatic electric field and ignoring loss due to ionization, we estimate the helium $1s2s$ $^1S_0$ multiphoton excitation rate with a femtosecond laser at 240 nm having a typical bandwidth of $\sim 5$ THz and obtain a bi-photon generation rate of $\sim 10^{11}$ $s^{-1}$ (see Supplementary Material and figure \ref{fig:3} (a)).     

An alternative scheme using a lambda-type transition between $1s^2$, $1s2p$, and $1s2s$ states could be used to achieve significant excitation. The energy levels of the latter two are 21.22 $eV$ and 20.62 $eV$ above the ground state, respectively. A two-step sequential excitation to first excite $1s^2\rightarrow 1s2p$ and then $1s2p\rightarrow 1s2s$ could be used. The oscillator strengths for one-photon excitation processes are $f_{a\rightarrow b}=2\Delta_{ba}|\langle b|\hat{\epsilon}_0\cdot\vec{r}|a\rangle|^2$, which gives $f_{1s^2\rightarrow 1s2p}=0.28$ and $f_{1s2p\rightarrow 1s2s}=-0.36$ for the two steps. To achieve this two-step sequential excitation, a high photon flux helium lamp source can be used in the first step to excite $1s2p$ and a 2059 nm laser can transfer population to the $1s2s$ state (see figure \ref{fig:3} (b)). The $\sim$ 1 GHz linewidth of the $1s2p$ state makes transitions to the $1s2s$ state using a broadband laser more feasible in comparison to direct multiphoton excitation. Currently available helium lamp sources are capable of generating $\sim 10^{15}$ photons $s^{-1}$. Using a high pressure helium target, nearly all of these photons could be absorbed to generate helium atoms in the $1s2p$ state. A high repetition-rate pulsed laser source at 2059 nm, could transfer nearly all these excited helium atoms to the $1s2s$ state. We estimate a bi-photon generation rate of $\sim 10^{13}$ $s^{-1}$ using this method (see Supplementary Material).

\begin{figure*}[t]
\includegraphics[width=0.99 \textwidth]{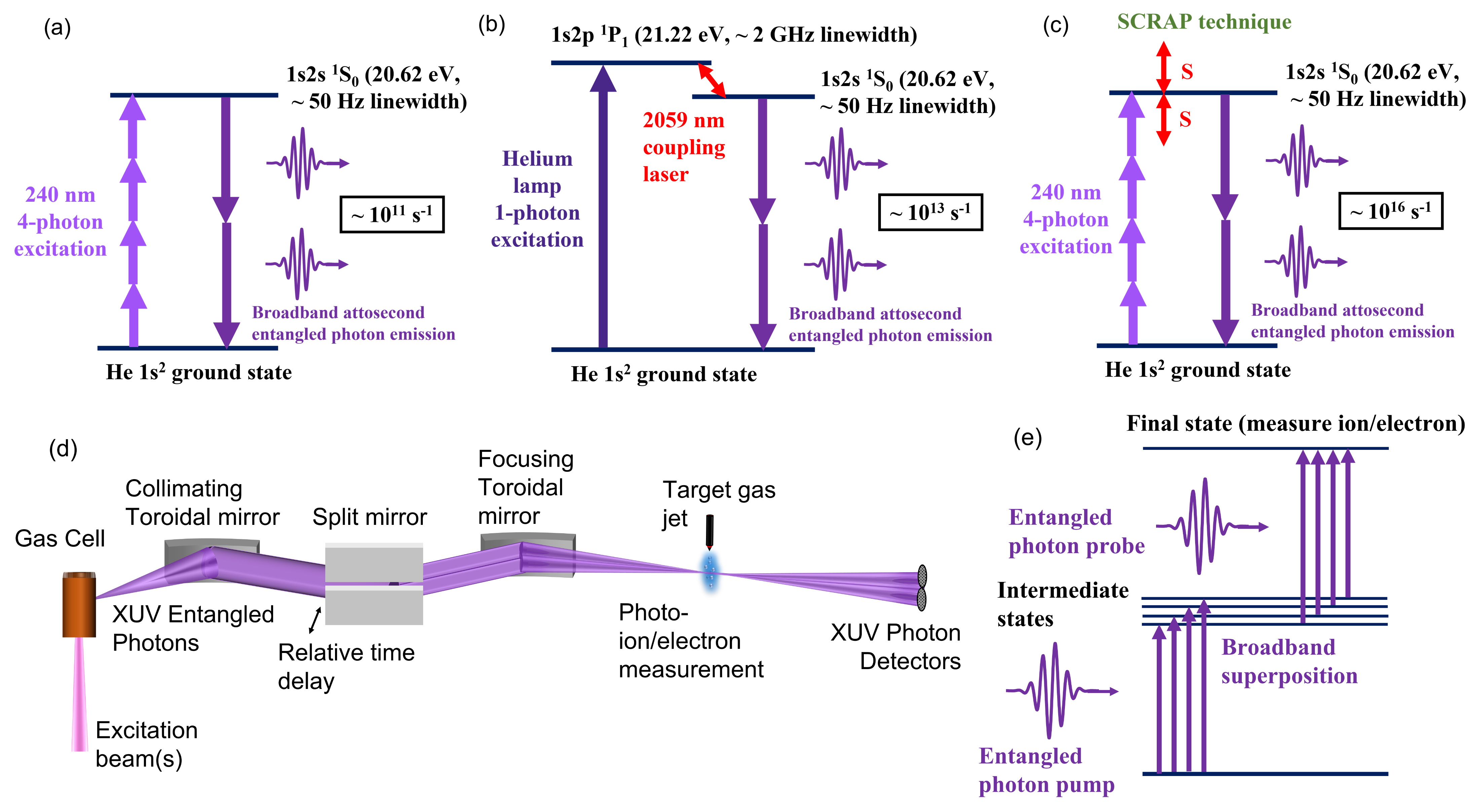}
\caption{\label{fig:3} (a) Generation of entangled bi-photons in the XUV via two-photon decay of the $1s2s$ $^1S_0$  state excited by four-photon excitation using a broad band 240 nm laser. (b) Two-step sequential excitation of the $1s2s$ state via the $1s2p$ state using a high photon flux helium lamp and a 2059 nm coupling laser. (c) The SCRAP technique to populate the $1s2s$ state using a multiphoton pump pulse and a Stark shifting pulse which enable rapid adiabatic passage and ionization suppression by LICS (LICS not shown). The estimated bi-photon generation rate is also shown for each scheme in (a) - (c). (d) Proposed experimental scheme to generate XUV entangled photons and utilize them in an attosecond pump-probe photoionization experiment. (e) An attosecond pump-probe photoionization scheme in molecules using entangled bi-photons.}
\end{figure*} 

Another alternative approach to achieve significant population of the $1s2s$ singlet metastable state is to use Stark-chirped rapid adiabatic passage (SCRAP), previously proposed to excite the 2s metastable state in a hydrogen atom \cite{yatsenko1999*,yatsenko2005}. In this technique, a pump pulse excites the metastable state via a multiphoton transition in the presence of a Stark pulse that Stark shifts the 1s2s state across the bandwidth of the pump pulse (see figure \ref{fig:3} (c)). The combined effect of the two pulses results in a Landau-Zener-type adiabatic passage that can significantly populate the 1s2s state. The SCRAP technique \cite{yatsenko2005}  can also suppress ionization loss by laser-induced continuum structure (LICS) \cite{knight1990,halfmann1998,yatsenko1999}. If we ignore ionization loss, for a typical femtosecond laser pulse-width of 50 fs with a bandwidth of 8.8 THz, rapid-adiabatic passage can excite nearly all atoms in the focal volume. When ionization loss is considered, since LICS can suppress ionization loss, it is reasonable to assume that $\sim 1\%$ of the atoms can be excited using SCRAP for every pair of pump and Stark pulses. With $\sim 10^{13}$ atoms in the focal volume corresponding to a 100 $\mu m$ spot size and 1 mm path length at 1 bar target pressure, this results in $\sim 10^{11}$ atoms excited per pulse. At a femtosecond pulse repetition rate of 100 kHz currently available, this results in an entangled bi-photon generation rate of $10^{16}$ $s^{-1}$ (see Supplementary Material). Among the three methods discussed here to excite helium to the singlet $1s2s$ state, the SCRAP method is expected to provide the highest excitation and hence the highest entangled bi-photon generation rate. 




The bi-photons from the decay of the $1s2s$ state are emitted in all directions with an approximate distribution given by $1+\cos^2(\theta)$ \cite{mokler2004}, where $\theta$ is the relative angle between the entangled photons. The photons that are emitted in a direction orthogonal to the excitation laser propagation direction can be collected within a large solid angle and sent along independent time-delayed paths towards a pump-probe target. Figure \ref{fig:3} (d) shows a schematic of a proposed experimental setup for generation of these entangled photons and their utilization in an attosecond pump-probe experiment. In this scheme, a grazing incidence toroidal mirror collimates the emitted photons which are then split into two halves using a grazing incidence split mirror that introduces a controllable time-delay between the two halves of the beam. Collecting bi-photons emitted along the same direction within a large solid angle (as opposed to those emitted in opposite directions), ensures that no time-smearing is introduced in the arrival times of the bi-photons. A $10\%$ collection solid angle will result in $1\%$ collection of bi-photons. The split beams are then focused using a second toroidal mirror onto the target gas jet. A pump-probe experiment with attosecond time resolution can be performed by measuring a photo-ion or photo-electron signal arising from the absorption of entangled bi-photons by an atom or molecule. Recent work on entangled two-photon absorption sets upper bounds on the enhancements in two-photon absorption cross section with entangled photons when no intermediate resonances are involved \cite{landes2021opex,landes2021prr}. Assuming a bi-photon rate of $\sim 10^{12}$ $s^{-1}$ at the pump-probe target and a two-photon cross section of $10^{-50}$ $cm^4s$, a pump-probe photoionization rate of $\sim 1000$ ions per second, which is well-above detection threshold of ion spectrometers, is expected. When intermediate resonances are involved, such as broad absorption resonances in molecules typically studied in attosecond experiments, this photoionization rate can be increased by a few orders of magnitude (see Supplementary Material). Further, measuring a pump-probe photoionization signal as opposed to a photon absorption signal as in previous two-photon absorption experiments allows detection of low absorption rates. Such entangled photon pump-probe experiments will extend the capabilities of attosecond science, where currently attosecond pulses from high-order harmonic generation \cite{gaumnitz2017} or free electron laser \cite{duris2020} sources are used.

The entangled photon generation scheme discussed here can be extended to the soft X-ray (SXR) regime using helium-like ions. Two-photon decay in helium-like ions has been well studied \cite{drake1969,mokler2004, bondy2020}. Similar to the 1s2s $^1S_0$ state of neutral helium atoms, the 1s2s $^1S_0$ states of helium-like ions such as N$^{5+}$, O$^{6+}$ and Ne$^{8+}$, predominantly decay by two-photon emission with a rate proportional to $Z^6$, where Z is the atomic number. The large energy difference between such excited states and the ground state of the ions, which can be in the range of several hundred to thousands of electron-volts, results in entangled photon correlation times of a few attoseconds to zeptoseconds. For example, the 1s2s $^1S_0$ state of Ne$^{8+}$ is located $\sim 915$ eV above the Ne$^{8+}$ ground state and this bandwidth corresponds to an entangled photon correlation time of $\sim 5$ attoseconds. The two-photon decay rate in this case is $\sim 1 \times 10^7$ s$^{-1}$ which is significantly larger than the corresponding rate for neutral helium atoms of $\sim 5 \times 10^1$ s$^{-1}$. Ne$^{8+}$ has been previously generated using strong femtosecond laser fields \cite{crane1995,chowdhury2003} as well as using strong femtosecond X-ray pulses from free electron lasers (FEL) \cite{young2010} both of which can potentially also create Ne$^{8+}$ in the $1s2s$ $^1S_0$ excited state. In one possible scheme, strong laser field ionization could generate Ne$^{8+}$ ions in the ground state and an FEL could excite them to the $1s2s$ $^1S_0$ state by two-photon excitation which then generate highly broadband entangled bi-photons at SXR energies. It has been previously demonstrated experimentally that the bandwidth required to generate few-attosecond pulses can be obtained from HHG using mid-infrared pulses \cite{popmintchev2012}. Further, it has been theoretically shown that zeptosecond pulses can be generated from HHG when suitable filters are used \cite{hernandez2013}. However, the shortest measured attosecond pulse is currently $43$ attoseconds \cite{gaumnitz2017}. Our approach of using entangled photons from two-photon decay of helium-like ions offers an alternative path for carrying out ultrafast measurements in these extreme regimes of a few-attoseconds to zeptoseconds. 


In conclusion, an unconventional approach is presented here for generating attosecond entangled bi-photons in the XUV and SXR regimes using two-photon decay in helium atoms and helium-like ions. Multiple alternative schemes can be used to excite the $1s2s$ $^1S_0$ metastable state in helium for which excitation rates have been estimated and an experimental scheme is suggested to collect and use the emitted XUV bi-photons in attosecond pump-probe experiments. The calculated photoionization rates indicate that attosecond pump-probe experiments with entangled photons are feasible. Potential extension of such metastable excitations to helium-like ions is additionally proposed, whereby SXR bi-photons can be generated with entanglement times in the few-attosecond range with the possibility of reaching the zeptosecond regime. This approach can open doors to using XUV/SXR entangled photons in quantum imaging and attosecond quantum spectroscopy of atomic, molecular and solid-state systems.

\begin{acknowledgments} 
The work of CHG and YW is supported by the U.S. Department
of Energy, Office of Science, Basic Energy Sciences,
under Award No. DE-SC0010545.
\end{acknowledgments}

\bibliography{entangled_refs1}

\end{document}



\title{Attosecond Entangled Photons from Two-Photon Decay of Metastable Atoms: A Source for Attosecond Experiments and Beyond: Supplementary Material}



\author{Yimeng Wang}
\email{wang3607@purdue.edu}
\author{Siddhant Pandey}%
\affiliation{Department of Physics and Astronomy, Purdue University, West Lafayette, Indiana 47907}
\author{Chris H. Greene}
\author{Niranjan Shivaram}
\email{niranjan@purdue.edu}
\affiliation{Department of Physics and Astronomy, Purdue University, West Lafayette, Indiana 47907}
\affiliation{Purdue Quantum Science and Engineering Institute, Purdue University, West Lafayette, Indiana 47907}%

\date{\today}

\begin{abstract}
This supplementary material provides details on the calculation of the geometry factor for the spheroid cavity, estimates for the entangled bi-photon generation rates in various schemes, and estimates for photoionization rates from entangled two-photon absorption in an attosecond pump-probe experiment. 
\end{abstract}

\maketitle

\section{\label{sec:spheroid} Geometry Factor For The Spheroid Cavity}
In this section, we will discuss the influence of the cavity geometry on the photoabsorption process, 
and evaluate the geometry factor $\Theta$ introduced in Eq. 7 in the main article. 
We start with summing over all the optical modes, to obtain an entangled-photon state: 
\begin{equation}\tag{S1}
    \sum_{s,s^{\prime}}=\frac{V^2}{(2\pi c)^6}\int \omega_s^2\omega_{s^{\prime}}^2 d\omega_s  d\omega_{s^{\prime}}
    d\Omega_{k_s} d\Omega_{k_{s^{\prime}}} \sum_{\hat{\epsilon}_{s,s^{\prime}}(\perp\hat{k}_{s,s^{\prime}})}
\end{equation}
where $\int d\Omega_{k_s}=\int \sin{\theta_s}d\theta_s d\phi_s$ integrates over the emission angle, and $\sum_{\hat{\epsilon}_{s,s^{\prime}}(\perp\hat{k}_{s,s^{\prime}})}$ sums the polarization vector basis. In a spherical cavity, the photon propagation vector is usually chosen to be (the mode index $s$ and $s^{\prime}$ for the two photons are neglected when we discuss only a single photon),
$$\hat{k}=(\sin{\theta}\cos{\phi},\sin{\theta}\sin{\phi},\cos{\theta})$$
When the photon is bounced back, its propagation vector become $\hat{k}^{\prime}=-\hat{k}$. The polarization vectors which are perpendicular to $\hat{k}$ can be expanded in into two basis,
\begin{equation*}
\begin{cases}
   \hat{\epsilon}^{(1)}=(-\sin{\phi},\cos{\phi},0)& \\
   \hat{\epsilon}^{(2)}=(-\cos{\theta}\cos{\phi},-\cos{\theta}\sin{\phi},\sin{\theta})  &  \\
\end{cases} 
\end{equation*}
 
However, in a spheroid cavity we need to reparametrize everything above. We assume the spheroid cavity has one major axis length $2a$ and two minor axes length $2b$ ($a\geq b$), the two foci are aligned along the major axis of the spheroid (which is $\hat{z}$ axis), and the distance between them is $2l$ ($l=\sqrt{a^2-b^2}$). The photon is emitted from one focus, and no matter what direction it propagates, it bounces back and passes through the other focus.
The propagation vector before and after the reflection can be parameterized as: 
\begin{equation*}
\begin{split}
   &\hat{k}=\frac{1}{L_+}(b \sin{\tilde{\theta}}\cos{\tilde{\phi}}, b\sin{\tilde{\theta}}\sin{\tilde{\phi}},l+a\cos{\tilde{\theta}}) \\ 
   &\hat{k}^{\prime}=\frac{1}{L_-}(-b\sin{\tilde{\theta}}\cos{\tilde{\phi}}, -b\sin{\tilde{\theta}}\sin{\tilde{\phi}},-l+a\cos{\tilde{\theta}})   \\
\end{split} 
\end{equation*}
where $L_{\pm}=\sqrt{b^2\sin^2{\tilde{\theta}}+(l\pm a\cos{\tilde{\theta}})^2}$. The polarization basis for $\hat{k}$ and $\hat{k}^{\prime}$ are, 
\begin{equation*}
\begin{cases}
 \hat{\epsilon}^{(1)}=(-\sin{\tilde{\phi}},\cos{\tilde{\phi}},0)& \\ \hat{\epsilon}^{(2)}=\frac{-1}{L_+}[(l+a\cos{\tilde{\theta}})\cos{\tilde{\phi}},(l+a\cos{\tilde{\theta}})\sin{\tilde{\phi}},-b\sin{\tilde{\theta}}]&    \\
\end{cases}
\end{equation*}
\begin{equation*}
\begin{cases}
 \hat{\epsilon}^{(1)\prime}=(-\sin{\tilde{\phi}},\cos{\tilde{\phi}},0)& \\ \hat{\epsilon}^{(2)\prime}=\frac{1}{L_-}[(-l+a\cos{\tilde{\theta}})\cos{\tilde{\phi}},(-l+a\cos{\tilde{\theta}})\sin{\tilde{\phi}},-b\sin{\tilde{\theta}}]&    \\
\end{cases}
\end{equation*}
$\hat{\epsilon}^{(1)}$ is perpendicular to the incident plane and doesn't change upon reflection, but $\hat{\epsilon}^{(2)}$ does: $\hat{\epsilon}^{(2)(\prime)}=\hat{k}^{(\prime)}\times\hat{\epsilon}^{(1)(\prime)}$. The angular integral is, 
\begin{equation}\tag{S2}
\begin{split}
    \int d\Omega_{k}&=\int \sin{\theta}d\theta d\phi=\int \left[\frac{a}{L_+} - \frac{l(l+a\cos{\tilde{\theta}})(a+l\cos{\tilde{\theta}})}{L_+^3}
    \right]\sin{\tilde{\theta}}d\tilde{\theta} d\tilde{\phi}
\end{split}
\end{equation}

We now consider the photon pair in modes $s$ and $s^{\prime}$, their propagation directions are random and independent with each other. However, constrains are set on their polarization directions. In the spontaneous decay from $e=1s2s$ $(^1S_0)$ to $g=1s^2$ state, only the isotropic part of the dipole operator can survive:  
\begin{align*}
    \langle g|\hat{\epsilon}_{s^{\prime}}\cdot\vec{r}|j \rangle\langle j|\hat{\epsilon}_s\cdot\vec{r}|e \rangle=\frac{\hat{\epsilon}_s\cdot\hat{\epsilon}_{s^{\prime}}}{3}\langle g|r|j \rangle\langle j|r|e \rangle
\end{align*}
In the absorption process, since the photons are not detectable inside the cavity, all the angular directions are integrated coherently:
\begin{equation}\tag{S3}
\label{cor-3}
\begin{split}
    &\sum_{i,j=1,2}\int d\Omega_{k_s} d\Omega_{k_{s^{\prime}}}
    \frac{\hat{\epsilon}_s^{(i)}\cdot\hat{\epsilon}_{s^{\prime}}^{(j)}}{3}
    (\hat{\epsilon}_{s}^{(i)\prime}\cdot\vec{r}_1)
    (\hat{\epsilon}_{s^{\prime}}^{(j)\prime}\cdot\vec{r}_2) \\
    &=\sum_{i,j=1,2}\sum_{k\mu}\frac{(-1)^{1+k+\mu}}{\sqrt{3}}
    [\vec{r}_1\, \vec{r}_2]^{(k)}_{-\mu} \int d\Omega_{k_s} d\Omega_{k_{s^{\prime}}}
    [\hat{\epsilon}_s^{(i)\prime} \hat{\epsilon}_{s^{\prime}}^{(j)\prime}]^{(k)}_{\mu}
    [\hat{\epsilon}_s^{(i)} \hat{\epsilon}_{s^{\prime}}^{(j)}]^{(0)}_0 \\
    &=\sum_{i,j=1,2}\int d\Omega_{k_s} d\Omega_{k_{s^{\prime}}}
    \frac{1}{9}(\hat{\epsilon}_s^{(i)}\cdot\hat{\epsilon}_{s^{\prime}}^{(j)})
    (\hat{\epsilon}_{s}^{(i)\prime}\cdot\hat{\epsilon}_{s^{\prime}}^{(j)\prime})
    (\vec{r}_1\cdot\vec{r}_2)\\
    &=\Theta \, \vec{r}_1\cdot\vec{r}_2
\end{split}
\end{equation}
Where $[...]^{(k)}_{\mu}$ is an rank-$k$ spherical tensor with component $\mu$($|\mu|\leq k$), as a result of the tensor product of two vectors \cite{Fano:1973}. Given the emitted-photon tensor have only rank-0 component, from the orthogonality of spherical tensors, we have $k=0,\mu=0$. So only rank-0 transition is allowed in the absorption process. This conclusion will no longer hold once the directions of the photons can be detected, i.e., by a recoil effect of the atoms. 
The geometry factor $\Theta$ is introduced to denote the polarization part of the integration. When the cavity is a prefect sphere, Eq. \ref{cor-3} gives $\Theta=\frac{64\pi^2}{27}$. The change of $\Theta$ versus the aspect ratio of the cavity can be found in Figure \ref{fig:S1},
from the range of $a=b$ to $a=148 b$. As the cavity becomes prolate spheroid shaped, $\Theta$ decrease with the aspect ratio $a/b$, and become stable at around $\Theta=\frac{8\pi^2}{9}$. 

\renewcommand{\thefigure}{S1}
\begin{figure}
\includegraphics[width=0.4 \textwidth]{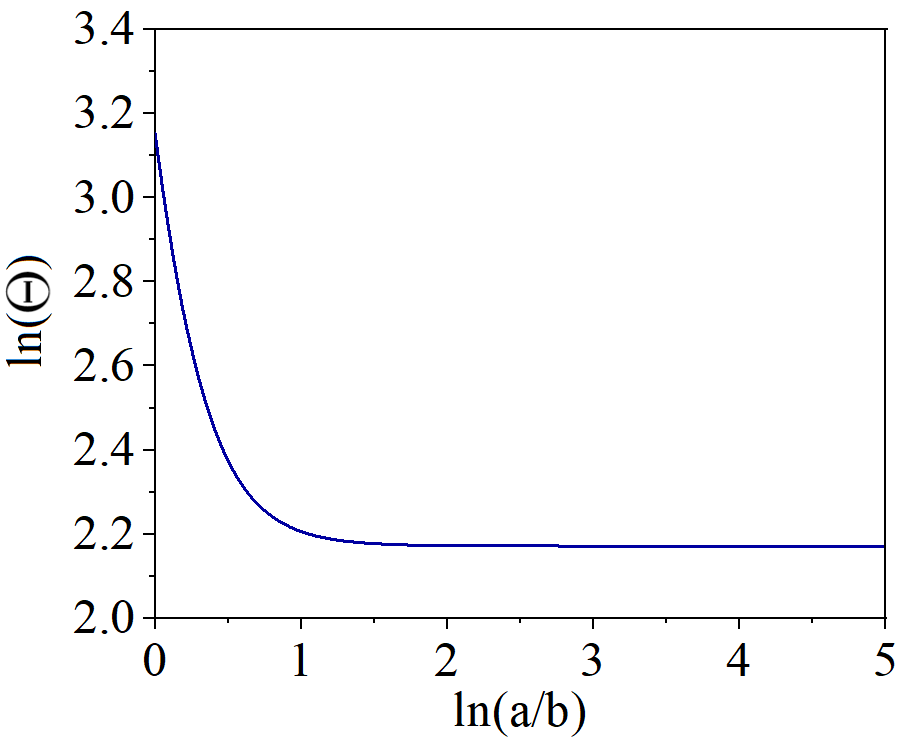}
\caption{ \label{fig:S1} A plot of geometry parameter $\Theta$ versus the aspect ratio $a/b$ in an log-log scale. When $a=b$, the cavity is a sphere and the geometry parameter obtained its maximum $\Theta=\frac{64\pi^2}{27}$. }
\end{figure}

\section{\label{sec:generation rate} Excitation rates for the helium $1s2s$ $^1S_0$ state}
In this section, we give estimates for the rate of bi-photon generation from the helium 1s2s state under realistic experimental conditions, for the schemes outlined in the main article. For a helium gas pressure of 1 bar, the number density $N$ $\sim 10^{19}$ atoms/cm\textsuperscript{3}. Such gas densities are easily achievable in static gas cells. In the following calculations, we will assume a pump focal spot diameter $d \sim 100$ um, and interaction length $L \sim 1$ mm.


\subsection{Four-photon excitation using 240 nm narrow-band and broad-band lasers}
The multiphoton absorption coefficient is related to the transition rate by the relation
\begin{equation}\tag{S4}
    \alpha^{(n)} = \frac{n\hbar\omega R^{(n)} N}{I^n}
\end{equation}
where $I$ is the intensity of the incident radiation, $R^{(n)}$ the n-photon transition rate, and $N$ is the number density in atoms/cm\textsuperscript{3}.
The variation in the intensity of radiation with distance is then given by
\begin{equation}\tag{S5}
    I(z) = I_0 e^{-\alpha^{(n)}z}
\end{equation}
for single-photon absorption, and
\begin{equation}\tag{S6}
    I(z) = \frac{I_0}{\Big(1 + (n-1)\alpha^{(n)} z I^{n-1}(0)\Big)^{\frac{1}{n-1}}}
\end{equation}
for multiphoton absorption. The per atom four-photon transition rate for the 1s2s state in helium at 240 nm can be calculated as:
\begin{equation}\tag{S7}
\label{SM_eqn:rate1}
    R^{(4)} = 2\pi \, \delta(\Delta_{eg}-4\omega_0) \bigg|\Big(\frac{e \mathcal{E}_0}{2\hbar}\Big)^4 D^{(4)}_{eg}\bigg|^2
\end{equation}
where $D^{(4)}_{eg}$=149 a.u. For a pump intensity $I_0 = 10^{14}$ W cm\textsuperscript{-2} (electric field strength of 0.053 a.u.) in the focus, the excitation rate is $\sim 3.4 \times 10^{-8} \, a.u.$, or $10^9 \, s^{-1}$ per atom. The absorption coefficient (in S.I. units) is given by
\begin{equation}\tag{S8}
    \alpha^{(4)} = \frac{4\hbar\omega_0 R^{(4)} N}{I^n} = 3.4\times10^{-46} \, W^{-3} cm^{5}
\end{equation}
The absorption fraction is given by
\begin{equation}\tag{S9}
    \frac{I_0 - I(L)}{I_0} = 1 - \frac{1}{(1 + 3 \alpha^{(4)} L I_0^3)^{1/3}} \sim 3.4\times10^{-5}
\end{equation}
The number of 240 nm photons per second in the focal volume for the given intensity (narrow-band continuous-wave laser) and focal spot size is $\sim 10^{28}$. This gives a bi-photon generation rate of $\sim 10^{22}$ per second.

The above calculation assumes a narrow-bandwidth ($< 50$ Hz) laser at 240 nm. To estimate the bi-photon generation rate for a broadband femtosecond 240 nm laser, we incoherently integrate over the different frequency components and write the $1s2s$ excitation rate (Eq. \ref{SM_eqn:rate1}) as
\begin{equation}\tag{S10}
\label{SM_eqn:rate_bb}
    R^{(4)}_{broad} = \int d\omega_0 \rho(\omega_0) 2\pi \, \delta(\Delta_{eg}-4\omega_0) \bigg|\Big(\frac{e \mathcal{E}_0}{2\hbar}\Big)^4 D^{(4)}_{eg}\bigg|^2
\end{equation}

For a broadband 240 nm laser with a bandwidth of 5 THz, an excitation rate and hence a bi-photon generation rate of $\sim 10^{11} $ $s^{-1}$ is obtained.

\subsection{Sequential excitation using helium lamp and 2059 nm laser}
We consider a two-step sequential excitation to first excite $1s^2\rightarrow 1s2p$ and then $1s2p\rightarrow 1s2s$. From our calculations, the oscillator strengths for one-photon excitation processes are $f_{1s^2\rightarrow 1s2p}=0.28$ and $f_{1s2p\rightarrow 1s2s}=-0.36$ for the two steps. The corresponding transition rates can be calculated using
\begin{equation}\tag{S11}
    R^{(1)} = \frac{\pi f}{\hbar\omega_0\mu} e^2 \mathcal{E}^2_0 \ \delta(\Delta_{eg}-\omega_0)
\end{equation}
where $\mu$ is the reduced mass, and the delta function is to be replaced by the lifetime of the excited state. Incoherent lamp sources that can generate $\sim 10^{15}$ photons/s, resonant with the $1s^2 \rightarrow 1s2p$ transition, and with a spot size of 100 um are commercially available (SPECS GmbH, $\mu SIRIUS$). The corresponding intensity for the lamp source is $34$ W cm\textsuperscript{-2}, and we assume an intensity of $10^{12}$ W cm\textsuperscript{-2} for the 2059 nm excitation laser. Assuming a 100 um spot size for both the excitation beams gives the transition rates $R_{1s^2\rightarrow 1s2p} \sim 3.7\times10^{9}$ and $R_{1s2p\rightarrow 1s2s} \sim 5\times10^{21}$ per second.

If we denote the lifetime of the 1s2p state by $\tau$, we can calculate the steady-state number density of excited atoms after the first excitation step using \cite{van2015}:
\begin{equation}\tag{S12}
    \frac{N_{exc}}{N} = \frac{1}{2} \Bigg(1 - \frac{1}{1+(2\sigma^{(1)}\tau/\hbar\omega)I}\Bigg) = \frac{1}{2} \Bigg(1 - \frac{1}{1+2 R^{(1)} \tau}\Bigg) \sim 0.47
\end{equation}
At a pressure of 1 bar, the focal volume contains $7.8\times10^{13}$ neutral atoms, 47\% of which are excited to the 1s2p state. Note that the number of 2059 nm photons for the given intensity and focal spot size is $\sim 8\times10^{28}$, which is much larger than the number of excited atoms in the focal volume. It is possible to saturate the $1s2p \rightarrow 1s2s$ transition and all the atoms in the $1s2p$ state could be promoted by the 2059 laser to the $1s2s$ state. This gives a bi-photon generation rate of $\sim 3.6\times10^{13}$ per second. Note that for a sufficiently strong pulsed 2059 nm laser at high repetition rates, both the number of photons per second in the focal volume and the excitation rate per atom will satisfy the above mentioned criteria, and thus, give a similar rate of bi-photon generation.

\subsection{Four-photon excitation using SCRAP}
We consider a scenario where rapid adiabatic passage (RAP) is used to transfer population from the $1s^2$ to the $1s2s$ state via a four-photon coupling. Neglecting ionization leakage from the $1s2s$ state, the transfer efficiency is limited only by non-adiabatic transitions between the adiabatic states, which will be calculated in this section. The single-photon Rabi frequency $\Omega_{eg} = \langle \vec{\mu_{eg}}\cdot\vec{E} \rangle /\hbar$ can be generalized for a four-photon transition as
\begin{equation}\tag{S13}
    \Omega_{eg} = \left(\frac{e\mathcal{E}_0}{2\hbar}\right)^4 D^{(4)}_{eg}
\end{equation}
for linearly polarized light. If we consider the ${1s^2, 1s2s}$ ($^1S_0$) subspace as a two-level system, the four-photon Rabi frequency at a 240 nm laser intensity of $10^{14}$ W cm\textsuperscript{-2} becomes
\begin{equation}\tag{S14}
    \Omega_{eg} = 7.35 \times 10^{-5} a.u. = 1.9\times 10^{13} s^{-1}
\end{equation}


Let us assume that a Stark pulse sweeps the transition energy across the entire bandwidth of a pump pulse of duration $\tau$ seconds and bandwidth $\delta$ Hz. We will further assume a static detuning $\delta/2$ for the four-photon pump pulse, such that the Rabi frequency $\Omega = \sqrt{\Omega^2_{eg}+\delta^2/4}$. The rate of leakage due to non-adiabatic transitions can be calculated using the Landau-Zener formula \cite{vutha2010}:
\begin{equation}\tag{S15}
    \Gamma(t) = \Omega^2 \frac{\gamma}{\Delta^2+\gamma^2/4}
\end{equation}
where $\gamma \sim \sqrt{\dot{\Delta}/4\pi}$. Assuming a linear Stark sweep, $\Delta(t) = t \, (\delta/\tau)$ Hz ($-\tau/2 \leq t \leq \tau/2$), and $\gamma = \sqrt{\delta/4\pi\tau}$. The corresponding transition rate is
\begin{equation}\tag{S16}
    \Gamma(t) = \frac{\Omega^2 \sqrt{\delta/4\pi\tau}}{t^2 \, (\delta/\tau)^2+\delta/16\pi\tau}
\end{equation}
For a pulse of duration $\tau$ = 50 fs and bandwidth $\delta$ = 8.8 THz, the probability of population transfer to the excited state, neglecting ionization leakage, is
\begin{equation}\tag{S17}
    1 - \exp\Bigg[\int_0^{\tau} \Gamma(t) dt\Bigg] = 1-\exp(-10.1) = 0.99996
\end{equation}

This shows that nearly all atoms in the focal volume can be excited. The SCRAP technique also involves ionization suppression by laser-induced continuum structure (LICS). Thus it is reasonable to assume that when ionization loss and LICS are considered, at least $1\%$ of the atoms in the focal volume are excited to the singlet $1s2s$ state for every pair of pump and Stark pulses. With $\sim 10^{13}$ atoms in the focal volume corresponding to a 100 $\mu m$ spot size and 1 mm path length at 1 bar target pressure, this results in $\sim 10^{11}$ atoms excited per pulse. At a femtosecond pulse repetition rate of 100 kHz currently available, this results in an entangled bi-photon generation rate of $10^{16}$ $s^{-1}$.  



\section{\label{sec:photoionization rate} Photoionization rate for entangled two-photon absorption in a molecule}

Recent work on entangled two-photon absorption \cite{landes2021opex, landes2021prr, raymer2021} sets upper bounds on the enhancements in two-photon absorption cross section with entangled photons when no intermediate resonances are involved. The entangled two-photon absorption cross-section (no intermediate resonances), $\sigma_e$ $(cm^2)$ is defined by a phenomenological two-photon absorption rate $R = \sigma_e I$ where $I$  $(cm^{-2}s^{-1})$ is the entangled photon flux density. $\sigma_e \sim (\sigma^{(2)} / A_eT_e)$ \cite{raymer2021} where $\sigma^{(2)}$ $(cm^4s)$ is the conventional two-photon cross section, $A_e$ is the entanglement area which is the focal spot size and $T_e$ is the entanglement time. Using a typical value of $10^{-50}$ $cm^4s$ for $\sigma^{(2)}$, $10^{-15}$ seconds for $T_e$ and $10^{-8}$ $cm^2$ for the spot size, $\sigma_e = 10^{-29}$ $cm^2$. For an entangled photon rate of $10^{12}$ per second, the two-photon absorption rate per molecule is $\sim$ $10^{-9}$ per second. With 1 bar target pressure, there are $10^{12}$ target molecules in the focal volume, resulting in $\sim$ 1000 ions generated per second. This rate is well above the detection threshold of a time-of-flight mass spectrometer. In molecular pump-probe experiments, strong intermediate resonances may be involved which further increase the ion rate by a factor of 10 to 100. This analysis shows that the XUV entangled photon pump-probe experiments proposed here are feasible.

\bibliography{entangled_refs1_SM}